\documentclass[journal]{IEEEtran}
\usepackage{amsmath,amsfonts}
\usepackage{algorithmic}
\usepackage{algorithm}
\usepackage{array}
\usepackage{textcomp}
\usepackage{stfloats}
\usepackage{url}
\usepackage{verbatim}
\usepackage{graphicx}
\usepackage[tight,footnotesize]{subfigure}

\usepackage{booktabs}

\usepackage{color} 

\usepackage{multicol} 
\usepackage{multirow}
\usepackage{colortbl}
\usepackage{xcolor}

\usepackage{bm}

\usepackage[colorlinks=true, linkcolor=blue, citecolor=blue]{hyperref}
\usepackage{cleveref}

\begin{document}

\title{Ray-Based Characterization of the AMPLE Model from 0.85 to 5 GHz}

\author{Lingyou~Zhou,~\emph{Graduate Student Member, IEEE,} Xin Dong, Kehai Qiu,~\emph{Graduate Student Member, IEEE,} Gang Yu,~\emph{Graduate Student Member, IEEE,} Jie Zhang,~\emph{Senior Member, IEEE,} and Jiliang Zhang,~\emph{Senior Member, IEEE}

	\vspace{-10pt}
	
	\thanks{
	    This work is supported in part by the Horizon Europe IPOSEE project (101086219), and is supported in part by the open research fund of National Mobile Communications Research Laboratory, Southeast University (2024D09).
	    
	    Corresponding Author: Jie Zhang (jie.zhang@sheffield.ac.uk).
		
		Lingyou Zhou, Xin Dong, and Jie Zhang are with the School of Electrical and Electronic Engineering, The University of Sheffield, Sheffield, S10 2TN, UK, and also with Ranplan Wireless Network Design Ltd., Cambridge, CB23 3UY, UK. 
		
     	Kehai Qiu is with the Department of Computer Science and Technology, University of Cambridge, Cambridge, CB3 0FD, UK, and also with the College of Engineering, Design and Physical Sciences, Brunel University of London, Uxbridge, UB8 3PH, UK.
     	
     	Gang Yu is with the School of Electrical and Electronic Engineering, The University of Sheffield, Sheffield, S10 2TN, UK, and also with 5G/6G Innovation Centre, University of Surrey, Surrey, GU2 7XH, UK.
     	
     	Jiliang Zhang is with the State Key Laboratory of Synthetical Automation for Process Industries and the College of Information Science and Engineering, Northeastern University, Shenyang, 110819, China, and also with the National Mobile Communications Research Laboratory, Southeast University, Nanjing, 210096, China.  
		
	}
}
\markboth{IEEE}%
{Shell \MakeLowercase{\textit{et al.}}: A Sample Article Using IEEEtran.cls for IEEE Journals}


\maketitle

\begin{abstract}
In this paper, we characterize the adaptive multiple path loss exponent (AMPLE) radio propagation model under urban macrocell (UMa) and urban microcell (UMi) scenarios from 0.85-5 GHz using Ranplan Professional. We first enhance the original AMPLE model by introducing an additional frequency coefficient to support path loss prediction across multiple carrier frequencies. By using measurement-validated Ranplan Professional simulator, we simulate four cities and validate the simulations for further path loss model characterization. Specifically, we extract the close-in (CI) model parameters from the simulations and compare them with parameters extracted from measurements in other works. Under the ray-based model characterization, we compare the AMPLE model with the 3rd Generation Partnership Project (3GPP) path loss model, the CI model, the alpha-beta-gamma (ABG) model, and those with simulation calibrations. In addition to standard performance metrics, we introduce the prediction-measurement difference error (PMDE) to assess overall prediction alignment with measurement, and mean simulation time per data point to evaluate model complexity. The results show that the AMPLE model outperforms existing models while maintaining similar model complexity.
	
\end{abstract}

\begin{IEEEkeywords}
Path loss, prediction, radio propagation, ray tracing.
\end{IEEEkeywords}

\vspace{-10pt}
\section{Introduction}

\IEEEPARstart{C}{hannel} modeling in sixth generation (6G) and beyond systems will need to address significantly more complex propagation scenarios across a wider range of frequencies \cite{6genv01,6genv02,chmdsvy}. As one of the key large-scale parameters (LSPs), path loss and radio propagation modeling play a crucial role in statistical channel models, as they constrain the overall channel modeling process, including the prediction of small-scale parameters (SSPs) and the construction of the channel matrix \cite{3gpp20, 3gpp03, 5gcmsig,qdga,metis,wnr2,mmmgc,gencm,imt20,cost}. In other words, to further enhance the statistical channel models based on current efforts, neglecting LSPs like path loss and focusing solely on SSPs will lead to half the result with twice the effort \cite{ample00}.   
 
Currently, deterministic models and empirical models are the two main types for radio propagation path loss predictions. For deterministic models, they are either based on the full-wave solutions such as finite-difference time domain (FDTD) \cite{fdtd97}, finite element method (FEM), method of moments (MOM) \cite{mom93}, and finite integration technique (FIT), or ray-based methods like ray-tracing and ray-launching \cite{rt01,gan23}. Since full-wave solutions are significantly more complex and primarily used for detailed channel modeling, deterministic models for path loss prediction typically refer to ray-based models only. Based on ray-optics, those ray-based models predict path loss using detailed environmental information and by simulating the complete propagation process. By applying Maxwell's equations with appropriate boundary conditions, they simulate the entire propagation channel, with path loss obtained as part of the modeling outcome. However, when path loss is the main outcome of interest, the requirements for detailed environmental data combined with the complexity of solving electromagnetic equations make these models difficult to deploy across diverse propagation scenarios \cite{gol05}.

In contrast, for simplicity, empirical models predict path loss using a small set of parameters extracted from measurements and/or ray-based simulations under different scenarios. Initially, empirical models such as the Okumura-Hata model, the European Cooperative for Scientific and Technical (EURO-COST) research COST-231 model \cite{rap96}, and the Stanford University Interim (SUI) model \cite{abh05} are proposed for their simplicity and fast computation. However, they only cover limited scenarios and often result in large prediction errors. To achieve both simplicity and accuracy in general path loss prediction, geometry-based stochastic models (GBSMs) construct path loss models extracted from measurements and/or ray-based simulations under different scenarios \cite{3gpp20, 3gpp03, 5gcmsig,qdga,metis,wnr2,mmmgc,gencm,imt20,cost}. Those models include the alpha-beta-gamma (ABG) model, the close-in (CI) free space reference distance model, and the CI model with a frequency-dependent path loss exponent (PLE) \cite{5gcmsig,rpppl}. Still, without redesigning the model structures, these models are limited by their lack of environmental considerations, which gradually leads to large prediction errors as environmental complexity increases.

\begin{figure*}[t]
	\centering
	\includegraphics[width = 1\linewidth]{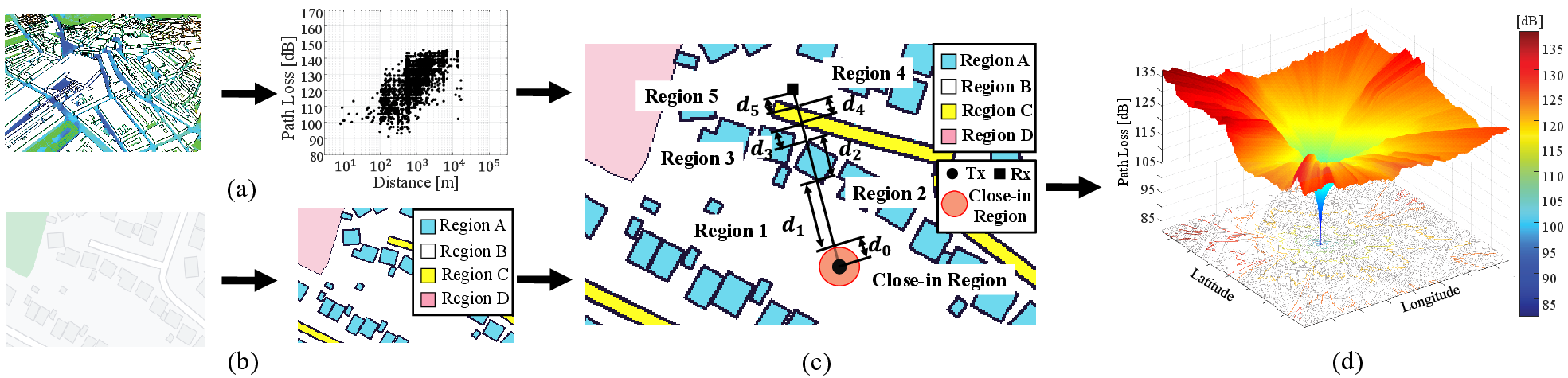}
	\caption{An example of the AMPLE model construction process \cite{ample00}. (a) Simulation/measurement collection and analysis. (b) Environment collection and classification. (c) Straight line construction. (d) Parameter regression and path loss prediction.} 
	\label{AMPCnstru}
	\vspace{-5pt}	
\end{figure*}

To cover propagation environments with low computational complexity, we previously proposed a fast adaptive multiple path loss exponent (AMPLE) model, and validated using both measurements \cite{ample00} and ray-based simulations \cite{ample01}. To enable practical deployment of the AMPLE model, in this paper, we characterize the AMPLE model under  urban macrocell (UMa) and urban microcell (UMi) scenarios from 0.85 to 5 GHz using Ranplan Professional. The main contributions of this paper are as follows.
\begin{itemize}
	\item{Based on our previous works \cite{ample00,ample01,ample02}, we enhance the AMPLE model by introducing an additional frequency coefficient to support path loss prediction across different carrier frequencies.}
	
	\item{We use Ranplan Professional, a measurement-validated ray-based simulator, to simulate and collect path loss data for fifth generation new radio (5GNR) 0.85 GHz, 2.1 GHz, and 5 GHz under UMa and UMi scenarios. Four cities are selected as propagation environments to set up the simulations, and the basic information including environments and propagation is based on the definitions of the typical UMa and UMi scenarios \cite{3gpp20,5gcmsig,rpppl}.}
	
	\item{We conduct a simple validation for the ray-based simulations, specifically under the field of path loss model characterization. That is, we extract the CI model parameters from the simulations and  for UMa line-of-sight (LOS), UMa non-line-of-sight (NLOS), and UMi NLOS environments. The comparison results show close alignment between the simulation-based and measurement-based CI model parameters.}
	
	\item{We characterize and validate the AMPLE model using Ranplan simulation data across the 0.85–5 GHz range for UMa LOS, UMa NLOS, UMi LOS, and UMi NLOS environments. The whole characterization and validation process is given, which can be considered as standard characterization process of the AMPLE model for future research.}
	
	\item{We compare the AMPLE model with the current path loss models in 3rd Generation Partnership Project (3GPP) \cite{3gpp20} and 5G Channel Model Simulation Group (5GCMSIG) \cite{5gcmsig} under the UMa/UMi scenarios with LOS/NLOS environments. Beyond general metrics such as root mean square error (RMSE) and mean absolute error (MAE), we propose prediction-measurement difference error (PMDE) to show overall alignment between predictions and measurements, and the mean simulation time per point to show model complexity. The overall results show that the AMPLE model outperforms the current empirical models by considering environments while maintains similar model complexity.}
\end{itemize}

The remainder of this paper is organized as follows. Section~\ref{Sec2} introduces the enhanced AMPLE model considering propagation carrier frequencies. Section~\ref{Sec3} presents the ray-tracing simulations conducted using the measurement-validated Ranplan Professional. In Section~\ref{Sec4}, the characterization process of the AMPLE model from 0.85-5 GHz is given, along with a performance comparison against other models. Finally, conclusions are drawn in Section \ref{Sec5}.



\section{The AMPLE Model}
\label{Sec2}
In this section, we give a brief background of the AMPLE model we previously proposed \cite{ample00,ample01,ample02}. And we also refine the AMPLE model by further considering propagation carrier frequency. An example of the model construction process is shown in Fig.~\ref{AMPCnstru}.

\subsection{Preliminary Information}
To construct the AMPLE model, both path loss and scenario information are necessary during the model construction stage \cite{ample00}. For path loss data, similar to other empirical path loss models \cite{3gpp20,3gpp03,5gcmsig,rpppl}, preliminary path loss data under the desired scenario are collected from ray-based simulations and/or measurements, as illustrated in Fig.~\ref{AMPCnstru}(a). After that, to combine the environment information and construct a site-specific model, the map information of the scenario is another preliminary model characterization condition. As shown in Fig.~\ref{AMPCnstru}(b), the AMPLE model utilizes the two-dimensional (2D) region map to cover environment information, of which it classifies the whole map into different region types. By using manual classification, map classification based on image processing, or artificial intelligence (AI)-based classification methods, the environment maps collected from satellite systems, digital map systems, and geographic information system (GIS) can be therefore transformed.

To combine path loss and map information, each type of region is assigned with a PLE, and a straight line between each transmitter-receiver (T-R) link is generated. The straight line records the intersected region PLE as well as the corresponding region length, which can be expressed as \cite{ample00,ample01}
\begin{equation}
	\label{LnMtx}
	\mathbf{S}_z =
	\begin{gathered}
		\begin{bmatrix}
			n_0   & n_1 & n_2  & \cdots  & n_{R_z}\\
			d_0 & d_1 & d_2  & \cdots  & d_{R_z} \\
		\end{bmatrix},
		\quad
	\end{gathered}
\end{equation}where $\mathbf{S}_z$ is the line matrix of the $z$th T-R link, $n_{R_z}$ is the PLE of the $R_z$th region, and $d_{R_z}$ is the corresponding region length. For common practice \cite{rap96,erc99}, we define regions within $d_0$ as the CI region, where $d_0$ is the CI distance \cite{ample00,ample01}. That is to say, regions within the CI distance are not counted (i.e., $n_{0}=0$). An example of the straight line and CI region is given in Fig.~\ref{AMPCnstru}(c).
 
\begin{figure}[t]
	\centering
	\includegraphics[width = 1\linewidth]{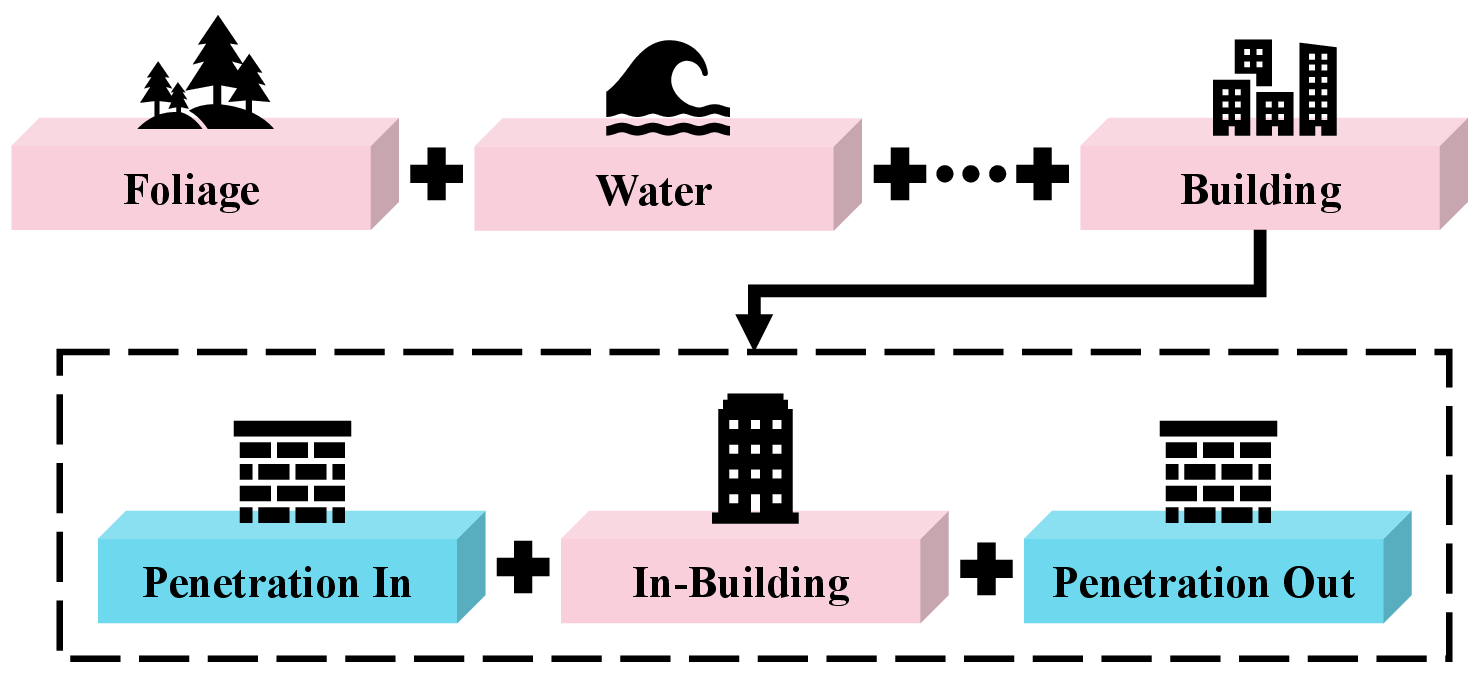}
	\caption{An example of building penetration loss.} 
	\label{VPen}
	\vspace{-5pt}	
\end{figure}

\subsection{The AMPLE Model}
By further considering the propagation carrier frequency as a factor, the decibel path loss of $z$th T-R link can be expressed as \cite{ample00,ample01,ample02}
\begin{equation}
\begin{aligned}
	\label{PLPEFs}
	\text{PL}_z\,[\text{dB}]= A + \sum_{r=1}^{R_z} 10 n_r \log_{10} &\left(\frac{\sum_{k=0}^{r} d_k}{\sum_{k=0}^{r-1} d_k}\right)+p_zX \\
	&+ 10 \gamma \log_{10}(f_z) +\Psi_{\sigma},
\end{aligned}	
\end{equation} 
where $A$, $R_z$, $n_{r}$, $d_k$, $p_z$, $X$, $\gamma$, $f_z$, and $\Psi_{\sigma}$ are characterized as follows.

\begin{itemize}	
	\item{\textbf{\emph{Intercept}} ($A$): Within the straight line, $A$ is the decibel path loss of the CI region.}
	
	\item{\textbf{\emph{Intersected Regions}} ($R_z$ $\&$ $n_r$): $R_z$ regions in total are intersected within the $z$th T-R straight line, and $n_r$ is the $r$th region PLE. Note that for different T-R pairs, $R_z$ may be different as well.} 
	
	\item{\textbf{\emph{Weighted Path Loss}} ($d_k$): The weighted path loss of the $r$th region is computed by the subtraction between the end point path loss of $r$th and $r-1$th region, respectively \cite{ample00,ample01,ample02}. Note that $d_k$ is the $k$th region length.}
	
	\item{\textbf{\emph{Penetration Loss}} ($p_z$ $\&$ $X$): We assume that there are two penetrations when the straight line intersects a building, each with an associated penetration loss $X$. For the $z$th T-R link, $p_z$ penetration losses are recorded along the straight line. Note that penetrations only exist when buildings are present (i.e., $p_z = 0$ if the scenario contains no buildings), as shown in Fig. \ref{VPen}.}
	
	\item{\textbf{\emph{Frequency}} ($\gamma$ $\&$ $f_z$): Based on the typical path loss characterization methods \cite{3gpp20, 3gpp03,5gcmsig,rpppl}, we add a coefficient $\gamma$ to indicate the dependence of path loss on frequency \cite{rpppl}, where $f_z$ denotes the carrier frequency of the $z$th link in GHz.}
	
	\item{\textbf{\emph{Shadowing}} ($\Psi_\sigma$): $\Psi_\sigma$ is a normally distributed shadowing term with $N[0,{\sigma}^{2}]$ under dB-scale.}
\end{itemize}
By combining terms with the same PLE, \eqref{PLPEFs} can be simplified as region-type-based, which can be expressed as
\begin{equation}
	\label{WtdPL_LkTrm}
	\text{PL}_z\,[\text{dB}] = A + \sum_{m=1}^{M} D_mn_m+p_zX+10 \gamma \log_{10}(f_z)+\Psi_{\sigma},
\end{equation}  
where $M$ is the number of region types within the environment (e.g., $M=4$ in Fig.~\ref{AMPCnstru}(c)), $n_m$ is the $m$th region type PLE, and $D_m$ is the corresponding coefficient extracted by combining terms of $n_m$. 

\begin{figure}[t]
	\centering
	\includegraphics[width = 1\linewidth]{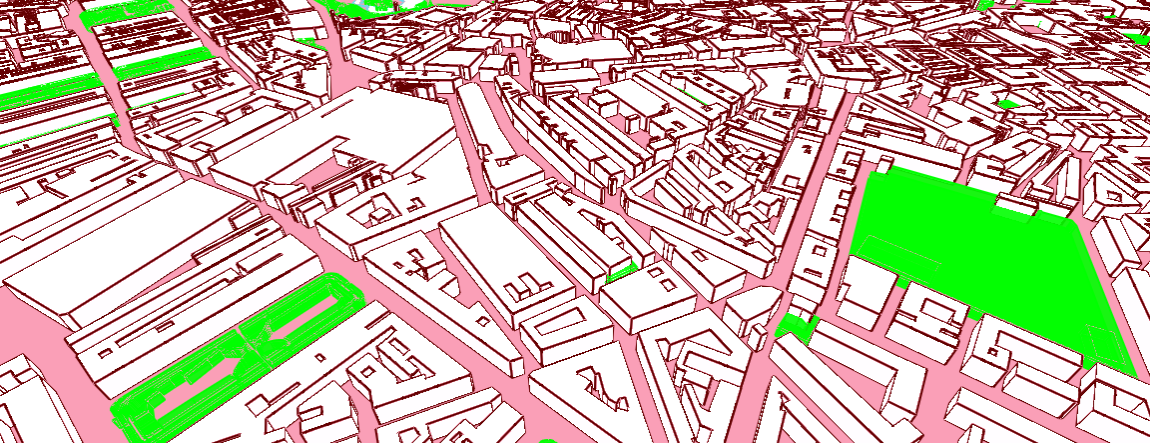}
	\caption{An example of the Ranplan Professional outdoor simulation in London.} 
	\label{SimExmp}
	\vspace{-10pt}	
\end{figure}

\section{Ranplan Professional Ray-Tracing Simulations}
\label{Sec3}
In this section, we first introduce the current usage of ray-based simulations and the measurement-validated Ranplan Professional simulator. We then provide details of simulations for the UMa and UMi scenarios using Ranplan Professional. Following that, we give a simple validation of the collected dataset by extracting model parameters such as PLE and comparing them to those obtained from measurements in \cite{rpppl}.

\subsection{Ranplan Professional Ray-Tracing Simulator}
With the development of ray-tracing technology, powerful ray-tracing simulators are used for hardware testbeds evaluations \cite{rtmea03,rtmea04,rtmea05}, essential 5G technology validations \cite{rtmea02}, channel model characterizations \cite{rtmea06}, and so on. These simulators have also been regarded as an alternative data collection method beyond hardware measurements due to their ability to collect larger data volume while with minor discrepancy with hardware measurement, and is called as “ray-tracing measurements" \cite{rtmea01}. In other words, for characterization of channel models, the database can be collected not only from hardware measurements, but also the ray-tracing simulations generated through reliable simulators. In this paper, we use Ranplan Professional \cite{rpln24}, a powerful commercial ray-tracing simulator, to characterize and validate the AMPLE model from 0.85 to 5 GHz. Validated by hardware measurements \cite{irla01, irla02}, Ranplan Professional has been used to characterize propagation models \cite{ample01, rpvli01} and support other research in wireless communications \cite{irla03,irla04,rpvli02,rpvli03}.


For the model characterization and validation, we focus on the UMa and UMi for outdoor scenario types. We consider the 5GNR system frequency bands at 0.85 GHz (5GNR band n5 with 10-MHz bandwidth), 2.1 GHz (5GNR band n1 with 10-MHz bandwidth), and 5 GHz (5GNR band n46 with 10-MHz bandwidth) across both scenario types, and we set up the simulation resolutions as 5 m. An example of the outdoor environment simulation is given in Fig.~\ref{SimExmp}, and the detailed propagation information for each scenario simulation is given as follows.

\begin{table*}[ht]
	\caption{Parameters of the CI Path Loss Model Extracted From Ranplan Professional Simulations and Measurements in \cite[Table I]{rpppl}. \# of Data Points Refers to the Number of Data Points After Distance Binning and Path Loss Thresholding.}
	\centering
	\renewcommand{\arraystretch}{1.3}
	\begin{tabular}{ccccccccc}
		\toprule
		\textbf{Scenario} & \textbf{Environment} & \textbf{Frequency [GHz]} &\textbf{Model} & \textbf{$\bm{d_0}$ [m]} & \textbf{Data Source } & \textbf{\# of Data Points} & \textbf{PLE} & \textbf{$\bm{\sigma}$ [dB]} \\ 
		\specialrule{.8pt}{0pt}{0pt}
		\multirow{4}{*}{UMa} & \multirow{2}{*}{LOS}  &  5GNR 2.1  & \multirow{2}{*}{CI} & \multirow{2}{*}{1}  &  Ranplan Professional &  12,283  &  2.26 &  5.06 \\ 
		&  & 2  &   & &  \cite[Table I]{rpppl}  & 253 & 2.00 &  1.70\\ \cline{2-9}
		& \multirow{2}{*}{NLOS} & 5GNR 2.1  & \multirow{2}{*}{CI} & \multirow{2}{*}{1} & Ranplan Professional & 211,996  & 2.92 & 10.08 \\ 
		&  & 2  & & &  \cite[Table I]{rpppl}  & 583 & 2.80 &  3.50\\ 
		\hline
		\multirow{2}{*}{UMi} & \multirow{2}{*}{NLOS}  & 5GNR 2.1  &\multirow{2}{*}{CI} & \multirow{2}{*}{1} & Ranplan Professional & 27,619  & 2.62 & 10.31 \\ 
		&  & 2.9  & & & \cite[Table I]{rpppl}  & 18 & 2.90 &  2.90\\ 
		\bottomrule
	\end{tabular}
	\label{RPPCmpr}
\end{table*}

\subsection{UMa Simulations in Sheffield and Barnsley}
For the UMa scenarios, we consider two outdoor simulations that are Sheffield and Barnsley in the UK. Both Sheffield and Barnsley are the typical European style cities that have medium population and building densities. We first import the city layout from EDINA’s Digimap Ordnance Survey \cite{osmap}, and we focus on the key environment regions that are general in real-world outdoor scenarios while have a significant impact on radio propagation. Those environment regions including \textit{buildings}, \textit{open space}, \textit{foliage}, and \textit{bodies of water} (lake, river, etc.). Note that, in order to the common use of the final model and the simplification of simulations, we here does not consider adding details into the outdoor scenario, such as road signs and other substances that are different in various places and are negligible in comparison to those key regions we covered \cite{ample00}. Within the simulations, regions of the same type (e.g., buildings) are assumed to have consistent material properties and heights. Following that, we set foliage with a height of 10 m, and the buildings are made of heavy concrete with a height of 20 m, which is close to the average height in the typical UMa scenarios \cite{3gpp20, rpppl}. For the scenario in Sheffield, it is bounded by latitudes and longitudes ranging from 53.36590854 to 53.39712604, and -1.5114482 to -1.46115222, respectively. In the meantime, the scenario in Barnsley is bounded by latitudes and longitudes ranging from 53.54123619 to 53.56927884, and -1.50333959 to -1.4537181, respectively.

Beyond that, the transmitters (Txs) in both scenarios are equipped with omnidirectional antennas transmitting at 5GNR frequencies of 0.85 GHz, 2.1 GHz, and 5 GHz as previously mentioned, with heights of 30 m \cite{rpppl}. For the exact positions, the Tx in Sheffield is positioned on the rooftop of a building at (53.381029, -1.4864733), and the Tx in Barnsley is positioned on the rooftop of a building at (53.5551977, -1.4789376). Along with that, the receiver (Rx) is positioned at a height of 1.5 m with the 5-meter resolution across the whole outdoor scenarios. The Tx power is set as 26 dBm, and both Tx and Rx are with 0 dBi antenna gains and with 0 dB cable loss. Based on the environments we set up, we collect a maximum of 823,923 raw data points from Sheffield and 682,803 raw data points from Barnsley across three carrier frequencies.


\subsection{UMi Simulations in London and Manchester}
For UMi scenarios, we simulate London and Manchester in the UK, which are two typical European style UMi cities. We focus on the center of two cities (i.e., London Soho and the center of Manchester), that are with high building densities and street canyons. The construction process explained in the UMa scenarios (i.e., Section~\ref{Sec3}-B) is used to set up the environments in two UMi cities (e.g., city layout and environment regions), except that the average building height is set to 10 m \cite{3gpp20, rpppl}. The scenario in London is bounded by latitudes and longitudes ranging from 51.48038 to 51.51303 and -0.16366 to -0.11336, respectively. Also, for scenario in Manchester, it is bounded by latitudes and longitude ranging from 53.47072 to 53.4978 and -2.27098 to -2.21759, respectively. 

For propagation information, the Txs in UMi scenarios have heights of 15 m (i.e., on the rooftop of a building), with the Tx in London positioned at (51.49474159, -0.14394048), and the Tx in Manchester positioned at (53.48710645, -2.24311856). Other propagation details, such as frequencies, Rx height, Tx power, and so on, are similar to those in the UMa scenarios. Due to the constraints of the UMi scenario, the overall signal coverage is significantly smaller than that of the full simulated environment in both cases. Finally, we collect 482,403 raw data points from London and 482,403 raw data points from Manchester across three carrier frequencies.


\subsection{Validation Based on CI Model PLE Extraction}

For path loss model characterization, a reliable dataset can produce similar model parameters (e.g., model PLE) compared to the PLEs obtained from measurements under similar propagation conditions recognized by professionals \cite{3gpp20, 5gcmsig, rpppl}. Under this way, to validate the simulation results for two scenario types, we compare the CI model parameters extracted from Ranplan Professional and measurements in \cite{rpppl}. In other words, after filtering the simulation data points, we first extract the CI model parameters by the same closed-form solutions as shown in \cite[Appendix Eq. (30) $\&$ (31)]{rpppl}, and we compared the CI values that are extracted by measurements in \cite[Table I]{rpppl}. Detailed process is given as follows. Note that for the CI model, it can be expressed as \cite{rpppl}
\begin{equation}
	\begin{aligned}
		\label{PLCI}
		\text{PL}_z^{\text{CI}}(f_z,d_z)\,[\text{dB}] &= A(f_z,d_0)\,[\text{dB}]
		\\
		& + 10n\log_{10}(\frac{d_z}{d_0}) + \Psi_{\sigma}^{\text{CI}}, \qquad     d_z \ge d_{0},
	\end{aligned}	
\end{equation}
with
\begin{equation}
	\label{FSPL}
	A(f_z,d_0)\,[\text{dB}] = 20\log_{10}\left(\frac{4\pi f_z d_0 \times 10^9}{c}\right),
	\vspace{-5pt}
\end{equation} 
where $A(f_z,d_0)$ is the free-space path loss in decibels with distance $d_0$ at carrier frequency $f_z$ in GHz, $n$ is the CI PLE, and $ \Psi_{\sigma}^{\text{CI}}$ is the shadowing term with $N[0,{\sigma}^{2}]$ under dB-scale.

To compare with the measurements around 2 GHz in \cite{rpppl}, we combine all data points for each scenario type, filter the interference points (e.g., data points beyond a path loss value, such as 150 dB; data points beyond a T-R distance, such as 1.5 km for the UMa scenario, etc.) and extract the path loss data at 5GNR 2.1 GHz. After that, we classify the data points into LOS and NLOS sets based on map information, which is similar to the classification method in \cite{rpppl}. By using closed-form solutions for CI PLE $n$ and standard deviation $\sigma$ \cite[Appendix Eq. (30) $\&$ (31)]{rpppl}, we extract those values under the UMa LOS, NLOS, and the UMi NLOS, and compared with those extracted from measurements in \cite{rpppl}, which are shown in Table~\ref{RPPCmpr}. 

Throughout Table~\ref{RPPCmpr}, the CI PLE $n$ extracted from Ranplan Professional are close to those in \cite{rpppl}, with a maximum difference below 0.3, whereas the differences of standard deviation $\sigma$ are large. The reasons of such differences including $n$ and $\sigma$ can be probably explained as: 
\begin{itemize}	
 	\item{ \textbf{Number of data points}: As shown in Table~\ref{RPPCmpr}, the number of data points exhibits significant discrepancies between Ranplan Professional simulations and measurements in \cite{rpppl}, with a maximum difference exceeding 1,500 fold. A larger number of data points is more likely to result in a higher variance in distribution, leading to an increased $\sigma$. Despite this, the PLE values of the two remain close, indicating a high similarity in the path loss trend between simulations and measurements.}
 	
 	\item{ \textbf{Frequency}: Differences between propagation carrier frequencies have a non-negligible impact on path loss, which further influences the model values. For example, as shown in the UMi NLOS case, the frequency difference (i.e., a 0.8-GHz difference) is one of the factors contributing to the PLE variation between simulation and measurement.}
 	
 	\item{ \textbf{Environment}: Even within the same scenario type, environmental differences may lead to minor variations of model values between different datasets.} 
 	
\end{itemize}

Overall, the simulation dataset of Ranplan Professional shows high similarity for the CI model PLEs in comparison to those in \cite[Table I]{rpppl}, and in this paper, we use these datasets to characterize the AMPLE model.


\section{Characterization of the AMPLE Model and Performance Results}
\label{Sec4}
\begin{figure}[t]
	\centering
	\includegraphics[width = 1\linewidth]{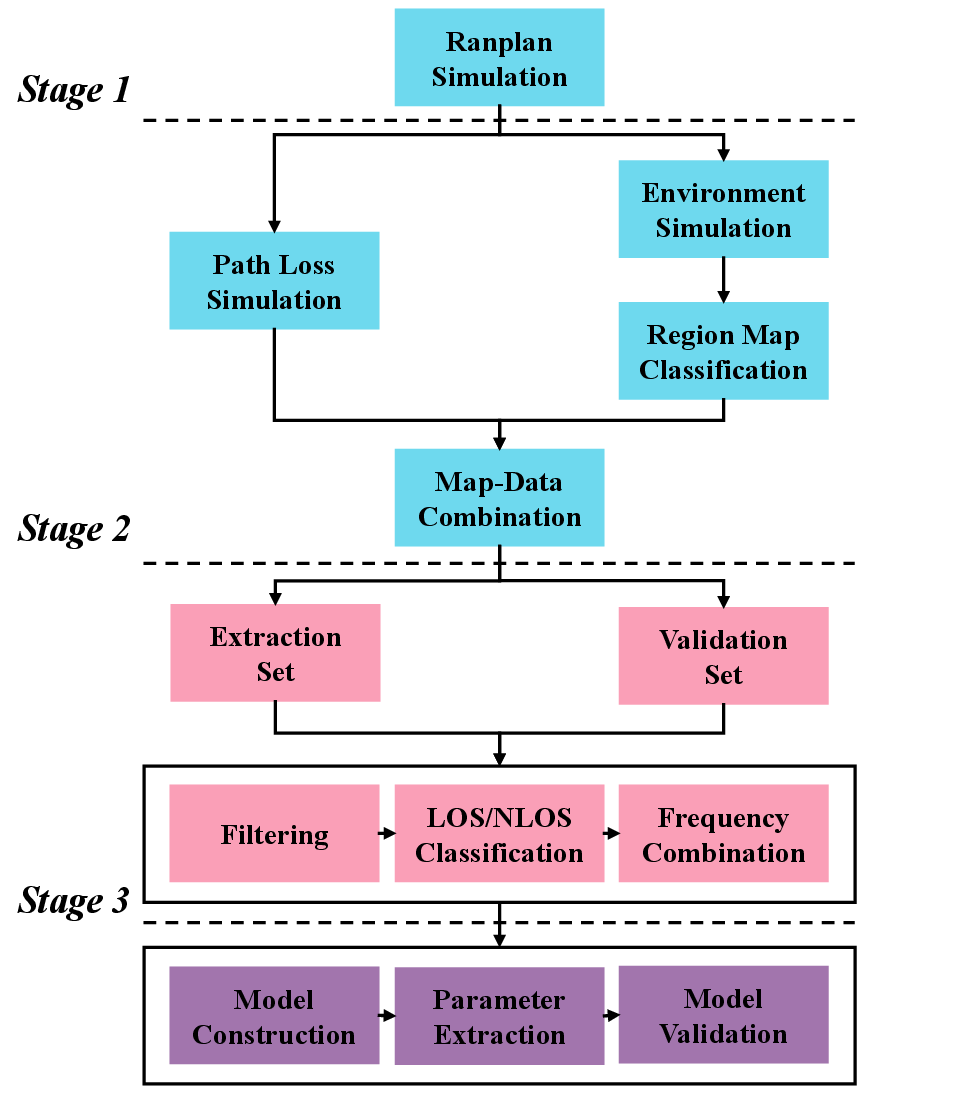}
	\caption{Flow chart of characterization process based on Ranplan Professional simulations, similar process could be considered for other types of datasets.} 
	\label{CProcs}
\end{figure}
\begin{figure*}[ht]
	\centering
	\includegraphics[width = 1\linewidth]{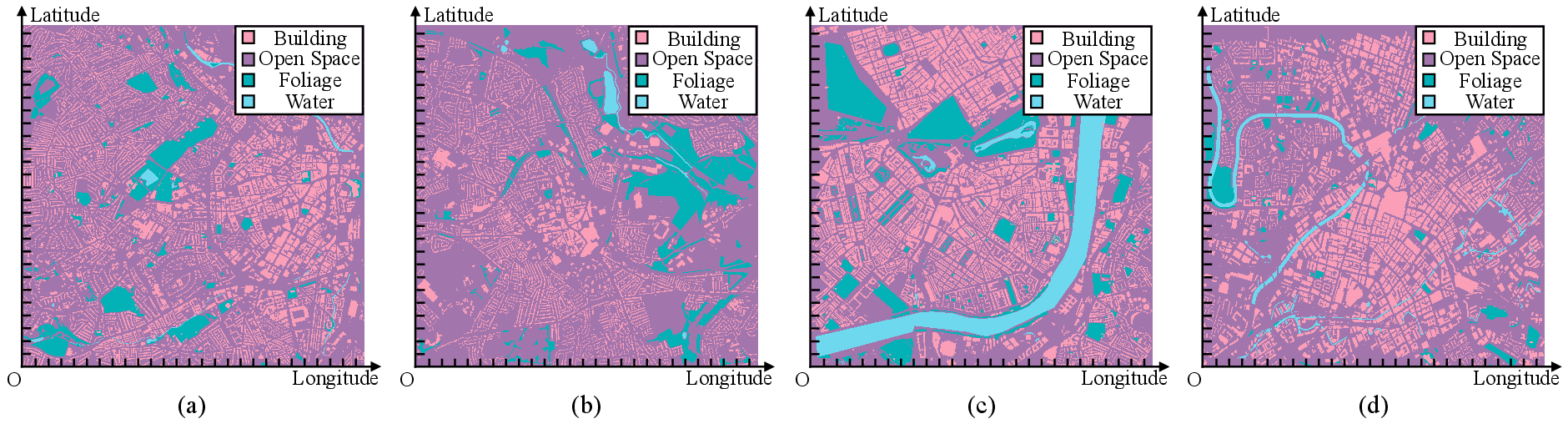}
	\caption{Region maps of (a) Sheffield, (b) Barnsley, (c) London, and (d) Manchester. Along with the classified region types and the geographic coordinates.} 
	\label{RgnMp}
\end{figure*}

In this section, we introduce the process of characterizing the AMPLE model based on Ranplan simulations which are presented in Section~\ref{Sec3}. Following that, we briefly introduce the compared models and the considered performance metrics including the two metrics we defined. Finally, we analyze the results and give the characterization parameters of the AMPLE model under the UMa and UMi scenarios from 0.85 to 5 GHz.
  
\subsection{Characterization Process of the AMPLE Model}

As shown in Fig.~\ref{CProcs}, the characterization process of the AMPLE model based on Ranplan Professional can be split into three stages, that is, region classification and map-data combination, simulation data processing, and model constructions.  

\textbf{Stage 1: Region Classification and Map-Data Combination.} Initially, Ranplan Professional contains both environment simulation and the corresponding path loss simulations. Following the environment types we set up as described in Section \ref{Sec3}-B, we first extract the region maps of those scenarios from Ranplan Professional. By extracting Ranplan Professional simulations as extensible markup language (XML) files, we collect the 2D region maps of four simulated cities, which cover region types including \textit{buildings}, \textit{open space}, \textit{foliage}, and \textit{bodies of water}. To map the path loss data with the generated region maps, we consider the geographical coordinates of the simulated scenarios. That is, by adding latitude/longitude to region maps, each path loss data can be therefore mapped into the region map. The region maps of four scenarios with coordinates are shown in Fig.~\ref{RgnMp}. 

\textbf{Stage 2: Data Processing.} To characterize and validate the AMPLE model, the datasets of four cities are split into extraction and validation sets, where an \textit{extraction set} refers to dataset used for model extraction, and a \textit{validation set} refers to dataset used for validating the model performance. We take Sheffield in the UMa and Manchester in the UMi as extraction set for parameter extraction, and Barnsley in the UMa and London in the UMi as validation set for model validation. Following the similar filtering and LOS/NLOS classification strategy described in Section \ref{Sec3}-D, we combine three frequencies together for model characterization, where the detailed information is listed in Table~\ref{InfoData}. 

\textbf{Stage 3: Model Extraction.} Considering the straight line method described in Section \ref{Sec2}, we construct the model format in \eqref{WtdPL_LkTrm} for each T-R pair. Based on the statistical properties of path loss and shadowing, we extract the desired model parameters including PLEs $n_m$, intercept $A$, penetration loss $X$, frequency coefficient $\gamma$, and the standard deviation $\sigma$. Detailed extraction process is described in Appendix~\ref{APP_AMPLE}.

\begin{table*}[b]
	\caption{Information of Datasets for Model Extraction and Performance Validation. \# of Data Points Refers to the Number of Data Points After Distance Binning and Path Loss Thresholding}
	\centering
	\renewcommand{\arraystretch}{1.3}
	\begin{tabular}{ccccccc}
		\toprule
		\textbf{Scenario} & \textbf{City} & \textbf{Set Type} &  \textbf{Environment}  & \textbf{Frequency Range [GHz]} &    \textbf{\# of Data Points} & \textbf{Distance Range [m]} \\ 
		\specialrule{.8pt}{0pt}{0pt}
		\multirow{4}{*}{UMa} & \multirow{2}{*}{Sheffield}  &  \multirow{2}{*}{Extraction}  & LOS & \multirow{2}{*}{0.85-5}  &  26,730 &  30-800   \\  
		
		&  &   &  NLOS &  & 373,570  & 35-1,500 \\ \cline{2-7}
		& \multirow{2}{*}{Barnsley} & \multirow{2}{*}{Validation}  & LOS & \multirow{2}{*}{0.85-5} & 36,331 &  31-800  \\ 
		&  &   & NLOS & &  391,134  & 31-1500 \\ \cline{1-7}
		
		\multirow{4}{*}{UMi} & \multirow{2}{*}{Manchester}  &  \multirow{2}{*}{Extraction}  & LOS & \multirow{2}{*}{0.85-5} &  4,431 &  18-200   \\
		
		&  &   &  NLOS &  & 36,436  & 18-600 \\ \cline{2-7}
		& \multirow{2}{*}{London} & \multirow{2}{*}{Validation}  & LOS & \multirow{2}{*}{0.85-5}& 3,396 &  28-200  \\ 
		&  &   & NLOS & &  37,109  & 28-600 \\ 
		\bottomrule
	\end{tabular}
	\label{InfoData}
\end{table*}





\subsection{Compared Models}

To validate the performance of the AMPLE model, we consider a comparison with other path loss models, including the 3GPP path loss models, the 5GCMSIG CI model (i.e., CI-5GCMSIG), the 5GCMSIG ABG model (i.e., ABG-5GCMSIG), the CI model with extraction-set-based calibration (i.e., CI-Calibrated), and the ABG model with extraction-set-based calibration (i.e., ABG-Calibrated). The 3GPP and 5GCMSIG path loss models are respectively shown in \cite[Table 7.4.1-1]{3gpp20} and \cite[Table6]{5gcmsig}, and for the CI-Calibrated and ABG-Calibrated models, the calibration process is given as follows.
  
As shown in \eqref{PLCI} and \eqref{FSPL}, the CI model can be expressed as \cite{5gcmsig, rpppl}
\begin{equation}
	\begin{aligned}
		\label{PLCI1m}
		\text{PL}_z^{\text{CI}}(f_z,d_z)\,[\text{dB}] = 20&\log_{10}\left(\frac{4\pi f_z \times 10^9}{c}\right)
		\\
		& + 10n\log_{10}\left(\frac{d_z}{1\,\text{m}}\right) + \Psi_{\sigma}^{\text{CI}}.
	\end{aligned}	
\end{equation}
Based on the extraction set for the UMa/UMi scenarios and LOS/NLOS environments, we construct the CI-Calibrated model using the same method employed to characterize the AMPLE model, and the details are given in the Appendix~\ref{APP_CI}. Note that the initial values for the extraction are based on \cite[Table 6]{5gcmsig}. 

Meanwhile, for the ABG model, it can be expressed as \cite{5gcmsig, rpppl}
\begin{equation}
	\begin{aligned}
		\label{PLABG1m}
		\text{PL}_z^{\text{ABG}}(f_z,d_z)\,[\text{dB}] &= 10\alpha\log_{10}\left(\frac{d_z}{1\,\text{m}}\right) \\ 
		&+\beta + 10\gamma^{\text{ABG}}\log_{10}\left(\frac{f_z}{1\,\text{GHz}}\right)+ \Psi_{\sigma}^{\text{ABG}},
	\end{aligned}	
\end{equation}
where $\alpha$ is the model PLE, $\beta$ is the intercept, and $\gamma^{\text{ABG}}$ is the frequency coefficient. Since \cite[Table 6]{5gcmsig} does not include the LOS environment for the UMa and UMi scenarios, the initial values for the  LOS and NLOS environments in model extraction are based on \cite[Table III]{rpppl} and \cite[Table 6]{5gcmsig}, respectively. Similar to the CI-Calibrated model, the details of calibration process are provided in Appendix~\ref{APP_ABG}.

\subsection{Performance Metrics}

For performance metrics, we consider the RMSE and MAE to assess the point-wise performance of model, and the average total hit ratio error (AHRE) to evaluate prediction quality of path loss \cite{owa01,ost03,ost10}. Furthermore, we define the PMDE to evaluate the overall alignment between prediction and measurement via statistical distributions, and the mean simulation time per data point to measure model complexity. 

\textbf{Point-Wise Evaluation.} For RMSE, it can be computed as
\begin{equation}
	\begin{aligned}
		\label{RMSE}
		\text{RMSE} = \sqrt{\sum_{z=1}^{Z}\frac{(\hat{l_z}-l_z)^2}{Z}},
	\end{aligned}	
\end{equation}
where $Z$ is the total number of data points, $\hat{l_z}$ is the $z$th predicted path loss, and $l_z$ is the $z$th simulated path loss. For MAE, it can be expressed as   
\begin{equation}
	\begin{aligned}
		\label{MAE}
		\text{MAE} = \sum_{z=1}^{Z}\frac{|\hat{l_z}-l_z|}{Z}.
	\end{aligned}	
\end{equation}

\textbf{Quality Evaluation.} For AHRE \cite{ost03}, it is defined based on the total hit rate (THR) \cite{owa01}. For THR, given a path loss threshold $L_T$, a prediction is considered correct if both the predicted path loss value $\hat{l_z}$ and the simulated/measured path loss value $l_z$ are either greater than, less than, or equal to $L_T$, regardless of the deviation between $\hat{l_z}$ and $l_z$ \cite{ost10}. The AHRE represents the average deviation from 100$\%$ THR and is defined as \cite{owa01,ost03,ost10}
\begin{equation}
	\begin{aligned}
		\label{AHRE}
		\text{AHRE} = \frac{1}{N_{L_T}} \sum_{L_T=L_{T,\text{min}}}^{L_{T,\text{max}}}(100\%-\text{THR}(L_T)),
	\end{aligned}	
\end{equation}
where $N_{L_T}$ is the number of THR points, $L_T$ is the path loss threshold. In this paper, we set $L_{T,\text{min}} = 80$ and $L_{T,\text{max}} = 100$ for LOS environments, and $L_{T,\text{min}} = 100$ and $L_{T,\text{max}} = 120$ for NLOS environments in both the UMa and UMi scenarios, as most path loss values are concentrated within these ranges. Additionally, this method is useful for evaluating the validity of a model in cases where coverage is defined solely by a threshold value. Note that a smaller AHRE indicates better model prediction accuracy \cite{owa01,ost03,ost10}.

\textbf{Overall Alignment Evaluation.} To evaluate the model performance from a more comprehensive perspective, we define the prediction-measurement distribution error (PMDE) as an overall alignment metric. We first use the Akaike information criterion (AIC) to determine the best-fit distributions for both the predicted and measured (or simulated) datasets \cite{as001}. We here consider the candidate distributions including normal, lognormal, gamma, Weibull, Rayleigh, Ricean, and chisquare. By constructing the probability density functions (PDFs) for the two datasets, we then compute PMDE as the integral of the absolute difference between the two PDFs. The PMDE can be given as
\begin{equation}
	\begin{aligned}
		\label{PMDE}
		\text{PMDE} = \int_x|f_{\text{p}}(x)-f_{\text{r}}(x)|dx,
	\end{aligned}	
\end{equation}
where  $f_{\text{p}}(x)$ and $f_{\text{r}}(x)$ are the PDFs of the predicted and measured (or simulated) datasets, respectively. Note that these two PDFs describe the overall distributions of the datasets across the environment, which is fundamentally different from point-based shadow fading that characterizes local variations at one fixed location \cite{rap96}. 

\begin{table*}[ht]
	\caption{Performance of Models under UMa Barnsley and UMi London from 0.85 to 5 GHz}
	\centering
	\renewcommand{\arraystretch}{1.2}
	\begin{tabular}{ccccccccc}
		\toprule
		\textbf{Scenario} & \textbf{Environment} & \textbf{Metric} & \textbf{3GPP} & \textbf{CI-5GCMSIG} & \textbf{CI-Calibrated} & \textbf{ABG-5GCMSIG}  & \textbf{ABG-Calibrated} & \textbf{AMPLE}\\
		\specialrule{.8pt}{0pt}{0pt}
		\multirow{10}{*}{\textbf{UMa}}& \multirow{5}{*}{\textbf{LOS}}  & \textbf{RMSE [dB]} & 5.95 & 9.52 & 5.07 & \multirow{5}{*}{N/A} & 4.77 & 4.67 \\
		&& \textbf{MAE [dB]} & 4.58 & 8.10 & 4.23 &  & 3.88 & 3.68 \\
		&& \textbf{AHRE [\%]} & 16.56 & 31.44 & 15.79 &  & 14.67 & 14.06 \\
		&& \textbf{PMDE} & 0.34 & 1.02 & 0.15 &  & 0.12 & 0.51 \\
		&& \textbf{\textbf{$\bm{t_p}$ [ns]} } & 110.51 & 13.12 & 12.91 &  & 12.46 & 9.00 \\ \cmidrule(lr){2-9}
		& \multirow{5}{*}{\textbf{NLOS}}  & \textbf{RMSE [dB]} & 15.34 & 11.05 & 11.43 & 10.75 & 11.35 & 9.43 \\
		&& \textbf{MAE [dB]} & 12.91 & 9.47 & 9.81 & 9.12 & 9.63 &  7.81\\
		&& \textbf{AHRE [\%]} & 21.21  & 19.05 & 19.57 & 18.01 & 18.87 & 15.41 \\
		&& \textbf{PMDE} & 1.02 & 0.77 & 0.80 & 0.63 & 0.71 & 0.56 \\ 
		&& \textbf{\textbf{$\bm{t_p}$ [ns]} } & 117.40 & 15.45 & 15.49 & 13.26 & 13.41 & 11.4 \\ 
		\specialrule{.8pt}{0pt}{0pt}
		\multirow{10}{*}{\textbf{UMi}}& \multirow{5}{*}{\textbf{LOS}}  & \textbf{RMSE [dB]} & 11.52 & 12.16 &  6.86 & \multirow{5}{*}{N/A} & 6.29 & 4.49 \\
		&& \textbf{MAE [dB]} & 10.71 & 11.39 & 5.78 &  & 5.02 & 3.65 \\
		&& \textbf{AHRE [\%]} & 37.95 & 39.89 & 23.15 &  & 20.39 & 14.45 \\
		&& \textbf{PMDE} & 1.33 & 1.37 & 0.83 &  & 0.97 & 0.26 \\ 
		&& \textbf{\textbf{$\bm{t_p}$ [ns]} } & 117.01 & 25.44 & 24.88 &  & 24.84 & 25.12 \\ \cmidrule(lr){2-9}
		& \multirow{5}{*}{\textbf{NLOS}}  & \textbf{RMSE [dB]} & 13.58 & 13.69 & 10.91 & 13.58 & 10.85 & 9.15 \\
		&& \textbf{MAE [dB]} & 11.54 & 11.71 & 8.17 & 11.54 & 7.93 & 6.54 \\
		&& \textbf{AHRE [\%]} & 35.51 & 36.31 & 26.15 & 35.51 & 25.45 & 20.35 \\
		&& \textbf{PMDE} & 0.93 & 1.00 & 0.69 & 0.93 & 0.73 & 0.56 \\ 
		&& \textbf{\textbf{$\bm{t_p}$ [ns]} } & 130.65 & 13.47 & 13.48 & 13.44 & 12.95 & 9.95 \\ 	
		\bottomrule
	\end{tabular}
	\label{PerfmMetrc}
\end{table*}
\begin{figure*}[h]
	\centering
	\subfigure[]{\includegraphics[width=.5\linewidth]{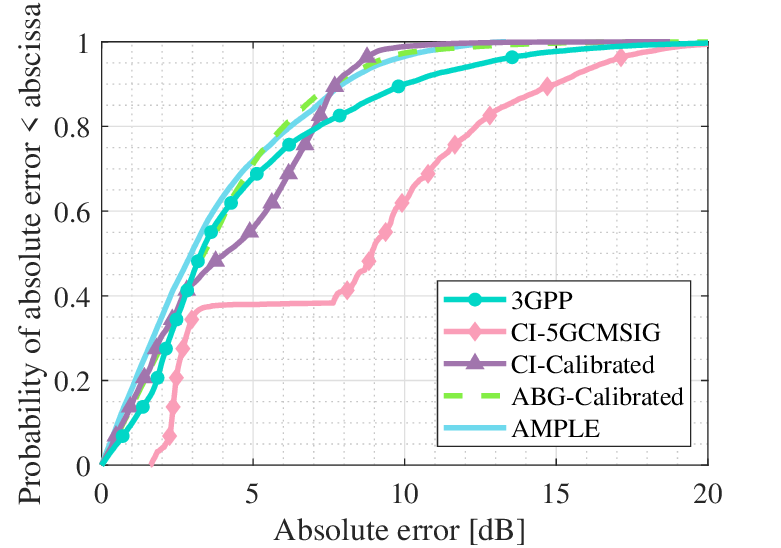}}%
	\subfigure[]{\includegraphics[width=.5\linewidth]{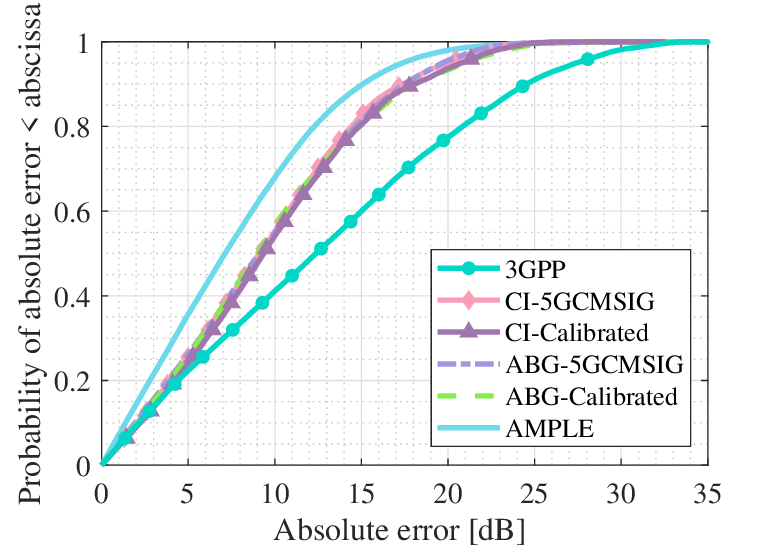}}
	\caption{CDF of absolute error between model predictions and Ranplan Professional simulations under the UMa Barnsley, with (a) the LOS environment, and (b) the NLOS environment.}
	\label{CDFUMa}		
	\subfigure[]{\includegraphics[width=.5\linewidth]{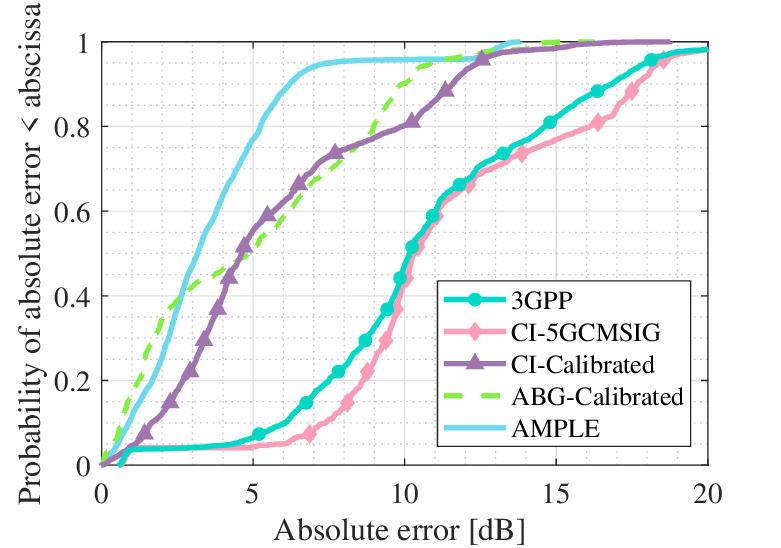}}%
	\subfigure[]{\includegraphics[width=.5\linewidth]{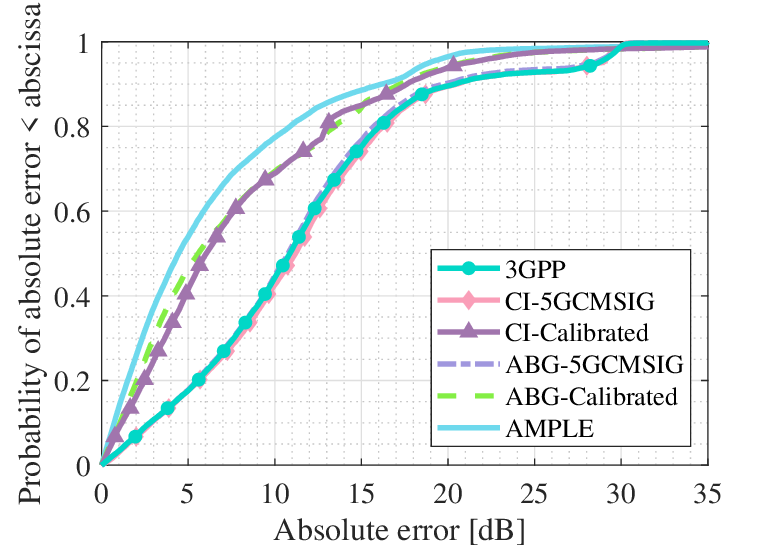}}
	\caption{CDF of absolute error between model predictions and Ranplan Professional simulations under the UMi London, with (a) the LOS environment, and (b) the NLOS environment.}
	\label{CDFUMi}	
	\hfil
\end{figure*}

The PMDE provides a statistically grounded way to evaluate the alignment between model-predicted and measured (or simulated) dataset distributions, going beyond point-wise errors and offering a distribution-level performance metric. Similar to RMSE and MAE, PMDE is a general-purpose metric that can also be applied beyond path loss modeling, wherever distributional similarity between predictions and reference data is of interest. A smaller PMDE indicates a better alignment between prediction and measurement/simulation. 

\textbf{Complexity Evaluation.} For the mean simulation time per data point, we run the models 1,000 times on the same dataset and compute the average simulation time for a single data point within one run. We define the mean simulation time per data point as  
\begin{equation}
	\begin{aligned}
		\label{MSTime}
		t_p = \frac{1}{CZ}\sum_{z=1}^{Z}\sum_{c=1}^{C}t_{z,c},
	\end{aligned}	
\end{equation}
where $t_p$ represents the mean simulation time per data point, $C$ is the total execution rounds (i.e., $C=1000$ in this case), and $t_{z,c}$ denotes the simulation time of $z$th data point in the $c$th execution (or run) of the model. We perform multiple runs instead of a single run and simulate the entire dataset rather than individual data points to mitigate fluctuations in computer performance. In this paper, we run the models and compute $t_p$ on a typical office computer (central processing unit (CPU): Intel (R) Core (TM) i5-10505 3.20 GHz; random-access memory (RAM): 16.0 GB 2133 MHz) with MATLAB-R2024b programming environment.

\begin{figure*}[ht]
	\centering
	\subfigure{\includegraphics[width=1\linewidth]{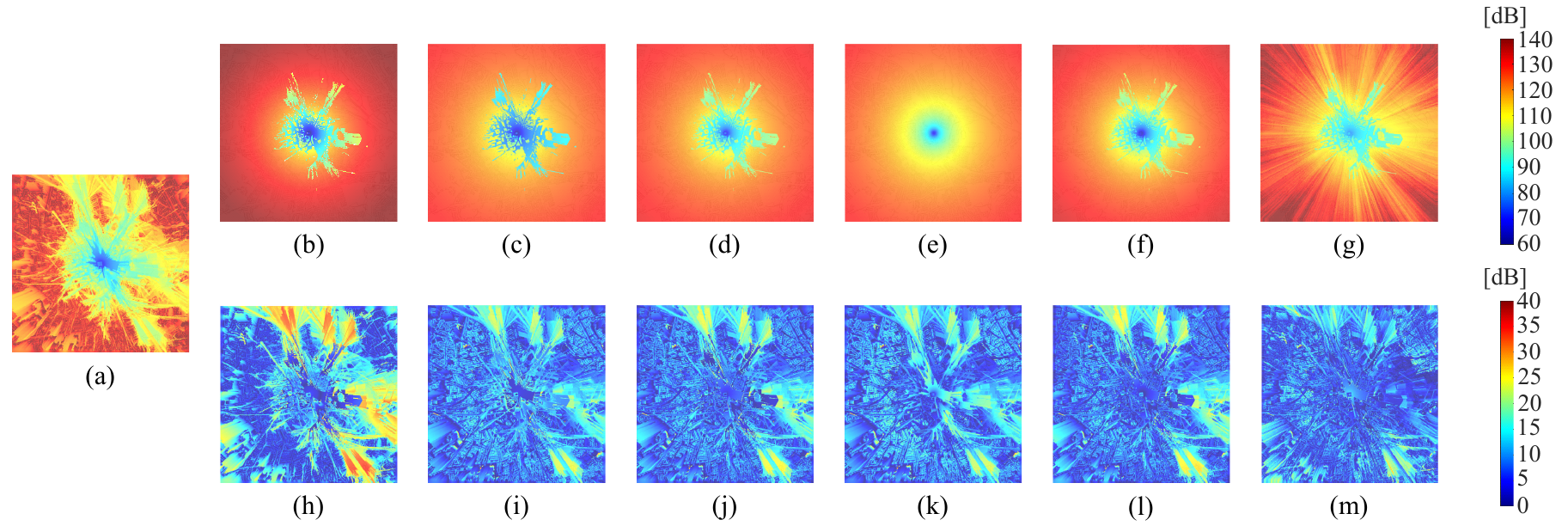}}%
	\caption{Predictions under the UMa Barnsley 0.85 GHz, which including: Heatmaps for (a) Ranplan Professional simulation, (b) 3GPP \cite[Table 7.4.1-1]{3gpp20}, (c) CI-5GCMSIG \cite[Table6]{5gcmsig}, (d) CI-Calibrated, (e) ABG-5GCMSIG \cite[Table6]{5gcmsig}, (f) ABG-Calibrated, and (g) AMPLE; and absolute error maps for (h) 3GPP, (i) CI-5GCMSIG, (j) CI-Calibrated, (k) ABG-5GCMSIG, (l) ABG-Calibrated, and (m) AMPLE.}
	\label{HMUMa}	
	\subfigure{\includegraphics[width=1\linewidth]{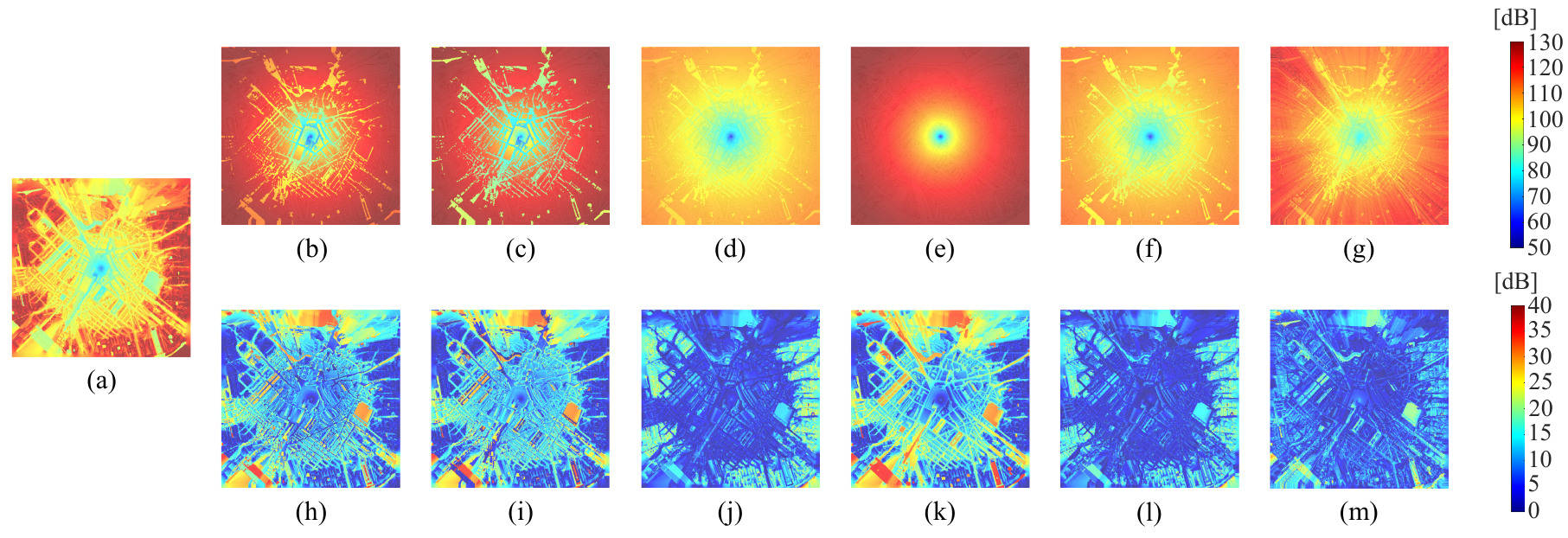}}%
	\caption{Predictions under the UMi London 0.85 GHz, which including: Heatmaps for (a) Ranplan Professional simulation, (b) 3GPP \cite[Table 7.4.1-1]{3gpp20}, (c) CI-5GCMSIG \cite[Table6]{5gcmsig}, (d) CI-Calibrated, (e) ABG-5GCMSIG \cite[Table6]{5gcmsig}, (f) ABG-Calibrated, and (g) AMPLE; and absolute error maps for (h) 3GPP, (i) CI-5GCMSIG, (j) CI-Calibrated, (k) ABG-5GCMSIG, (l) ABG-Calibrated, and (m) AMPLE.}
	\label{HMUMi}
	\hfil
\end{figure*}

\subsection{Results and Analysis}

As shown in Table~\ref{PerfmMetrc}, the performance of the models is evaluated using the metrics introduced in Section~\ref{Sec4}-C. Generally, the AMPLE model outperforms the current empirical path loss models used in GBSMs \cite{3gpp20,5gcmsig}, while maintaining the same level of simulation time, that is, the same model complexity. Note that when computing the simulation time $t_p$ in \eqref{MSTime}, similar to other empirical models, we consider the map information within the AMPLE model as pre-information (i.e., similar to simulated/measured path loss data) during model construction (as shown in Fig.~\ref{CProcs} and explained in Section~\ref{Sec4}-A). Moreover, we draw the cumulative distribution functions (CDFs) to visualize the absolute error between the model predictions and the simulation results under two scenarios (i.e., $|\hat{l_z}-l_z|$), which are shown in Fig.~\ref{CDFUMa} and  Fig.~\ref{CDFUMi}. 

For the UMa scenario under the LOS environment, the AMPLE model performs similarly to other models. This can be seen in both Table~\ref{PerfmMetrc} and Fig.~\ref{CDFUMa}(a), where the performance metrics in Table~\ref{PerfmMetrc} and the CDF plot in Fig.~\ref{CDFUMa}(a) show that the AMPLE model has similar performance compared to other models, such as the ABG-Calibrated model and the CI-Calibrated model. This is because the LOS case involves a simple environment where transmission experiences minimal environmental impact. In such a situation, considering environmental factors has limited contribution to path loss prediction. However, for the UMa scenario under the NLOS environment, environmental factors such as buildings and foliage cannot be ignored, making their consideration significant in path loss prediction \cite{ample00,ample01}. As shown in Table~\ref{PerfmMetrc}, the AMPLE model outperforms other models by considering main environmental factors, which provide more information during path loss predictions, and similar results can also be observed in Fig.~\ref{CDFUMa}(b). Meanwhile, in the more concentrated and complex UMi street canyon scenario, transmission is subject to greater environmental impacts, making environmental considerations even more crucial for path loss predictions. As a result, the AMPLE model outperforms other models, as shown in Table~\ref{PerfmMetrc} and Fig.~\ref{CDFUMi}.

Besides, to visually illustrate the performance of the AMPLE model, we present heatmaps and absolute error maps of the models under 0.85 GHz, as shown in Fig.~\ref{HMUMa} and Fig.~\ref{HMUMi} for the UMa and UMi scenarios, respectively. The absolute error map is computed based on the absolute error between the model predictions and Ranplan simulations. For the heatmaps in both scenarios (i.e., Fig.~\ref{HMUMa}(b)-(g) and Fig.~\ref{HMUMi}(b)-(g)), the predictions follow a radiation-like, distance-based trend under the LOS and NLOS cases, except that the AMPLE model predicts path loss based on both LOS/NLOS environments and environmental factors.

For the error maps in the UMa scenario (as shown in Fig.~\ref{HMUMa}(h)-(m)), prediction accuracy is largely influenced by the environment. Taking the 3GPP model as an example (Fig.~\ref{HMUMa}(h)), prediction errors are acceptable in the LOS environment but abruptly increase as the environment gradually transitions to the NLOS case. Even with different parameters for the NLOS environment \cite[Table 7.4.1-1]{3gpp20}, the 3GPP model still has a mean error of around 12 dB. This is similar to the CI-5GCMSIG model (Fig.~\ref{HMUMa}(i)) and the ABG-5GCMSIG model (Fig.~\ref{HMUMa}(k)), both of which perform slightly better than the 3GPP model in the NLOS environment. We previously consider that this may be caused by the differences in model construction datasets, so we use Ranplan simulation data to calibrate both the CI model and the ABG model (details are provided in Section~\ref{Sec4}-B). However, as shown in Fig.~\ref{HMUMa}(j) (the CI-Calibrated model) and Fig.~\ref{HMUMa}(l) (the ABG-Calibrated model), the results indicate that calibration improves predictions only in LOS cases, while a gap remains in the NLOS case. Most errors are similar to those in the CI-5GCMSIG model and the ABG-5GCMSIG model, with a mean error of approximately 9 dB. Following that, by considering environmental factors, the AMPLE model (Fig.~\ref{HMUMa}(m)) mitigates prediction errors in the NLOS case, with a mean error of around 8 dB.

For the error maps in the UMi scenario (as shown in Fig.~\ref{HMUMi}(h)-(m)), prediction accuracy is influenced by both the environment and the dataset. The same environmental impact is observed in this scenario, where the 3GPP model (Fig.~\ref{HMUMi}(h)), the CI-5GCMSIG model (Fig.~\ref{HMUMi}(i)), and the ABG-5GCMSIG model (Fig.~\ref{HMUMi}(k)) show significant prediction errors under the NLOS environment, with a mean error of approximately 12 dB. After calibration, the CI-Calibrated model (Fig.~\ref{HMUMi}(j)) and the ABG-Calibrated model (Fig.~\ref{HMUMi}(l)) show lower errors in the LOS environment but still have considerable errors in the NLOS environment, with a mean error of around 8 dB. Eventually, the environmental influences on predictions, especially in the NLOS environment, are effectively mitigated by the AMPLE model (Fig.~\ref{HMUMi}(m)), with a mean error of around 7 dB.


\subsection{The AMPLE model from 0.85 to 5 GHz}

\begin{table}[t]
	\centering
	\caption{Parameters of the AMPLE Model for UMa and UMi from 0.85 to 5 GHz}
	\renewcommand{\arraystretch}{1.5}
	\begin{tabular}{ccccc}
	  \toprule
	 \multirow{2}{*}{\textbf{Model Parameter}}  & \multicolumn{2}{c}{\textbf{UMa}} & \multicolumn{2}{c}{\textbf{UMi}} \\ \cmidrule(lr){2-3} \cmidrule(lr){4-5}
       & \textbf{LOS} & \textbf{NLOS}              & \textbf{LOS} & \textbf{NLOS} \\
      \specialrule{.8pt}{0pt}{0pt} 
      Intercept $A$ & 59.86 & 59.79 & 55.19 & 55.20\\
	  In-Building  $n_1$ & 1.35 & 1.80 & 1.59 & 1.78 \\
	  Open Space $n_2$ & 1.14 & 1.64 & 1.46 & 1.89 \\	
	  Foliage $n_3$   & 2.59 & 2.71 & 2.70 & 2.70 \\
	  Bodies of Water $n_4$ & 1.79 & 1.93 & 1.80 & 1.80 \\
	  Penetration $X$ & 0.09 & 0.28 & 0.18 & 0.17 \\
	  Frequency Coefficient $\gamma$ & 0.92 &  1.94& 1.97 & 1.98 \\
	  Shadow Fading std. $\sigma$ & 5.40 &  9.53& 8.01 & 8.00 \\
	   Default Values & \multicolumn{2}{m{2cm}}{ \raggedright Frequency from 0.85-5 GHz, $h_{\text{BS}}=30$ m, $h_{\text{UT}}=1.5$ m}& \multicolumn{2}{m{2cm}}{\raggedright Frequency from 0.85-5 GHz, $h_{\text{BS}}=15$ m, $h_{\text{UT}}=1.5$ m} \\
		\bottomrule
	\end{tabular}
	\label{AMP_Pmtr}
\end{table}

Table~\ref{AMP_Pmtr} lists the parameters of the AMPLE model for the UMa and UMi scenarios. With different antenna heights (i.e., UMa $h_{\text{BS}}=30$ m, and UMi $h_{\text{BS}}=15$ m), the intercept $A$ in the UMa scenario is slightly larger than in the UMi scenario. Beyond that, since most environments consist of buildings and open space, the PLE values of these two region types (i.e., $n_1$ and $n_2$) along with the building penetration loss $X$ largely influence the performance of the AMPLE model, resulting in distinct differences across environments/scenarios. In contrast, the PLEs of foliage and water (i.e., $n_3$ and $n_4$) remain similar across four cases. The frequency coefficient reflects the impact of frequency on path loss, where only the UMa LOS environment shows a weaker frequency dependence. This is acceptable as frequency coefficients may take relatively small or even negative values, indicating a minor influence on path loss \cite[Table III]{rpppl}.

In the meantime, two points require further clarification: (1) why the AMPLE model in the LOS environments still cover those environmental factors instead of modeling open space only,  despite the typical definition of LOS; and (2) why the open space regions in the LOS environments of both scenarios have relatively lower PLE values. Regarding the first point, we classify data points based on both the map layout and a path loss threshold. That is, the LOS category includes not only direct line-of-sight links but also those with first-order reflections, which can still result in low path loss values \cite{rtmea06}. Furthermore, according to the straight line mechanism described in Section  \ref{Sec2}-A, all regions intersected by the straight line are considered to contribute partially to the total path loss of the link. Therefore, even in the LOS environments, the influence of environmental factors remains present in both scenarios.

Moreover, the low PLE value for open space within the LOS environment is also attributed to the model's mechanism. Since the Txs are placed on buildings and the environmental information is based on 2D maps, links without obstructions are consistently misrecognized as penetrating through at least one building (i.e., the one located beneath the Tx) before reaching the Rxs. In other words, the impact of open space tends to be underestimated because the model mistakenly attributes it to buildings — particularly in short-range LOS cases. Under these conditions, most of the open space influence is incorporated into the characterization of buildings, causing the PLE $n_2$ to remain low at 1.14 for the UMa LOS and 1.46 for the UMi LOS. Instead, due to the more complex region intersections in the NLOS environments, this type of misclassification has less impact, as reflected in the more realistic parameter values observed in the NLOS cases for both scenarios \cite{3gpp20,5gcmsig}. While this misrecognition affects the PLE values in the LOS environments, it still results in good path loss prediction performance (as described in Section~\ref{Sec4}-D). Therefore, we consider addressing this issue as part of future work to further enhance the AMPLE model, particularly in the LOS environments.

\section{Conclusions}
\label{Sec5}

We have characterized the AMPLE model under UMa and UMi scenarios from 0.85-5 GHz. By using Ranplan Professional, we simulated four cites to characterize and validate the AMPLE model. The ray-based simulations at 2.1 GHz are validated with measurements by extracting the CI model parameters and compared with those extracted from measurements in \cite[Table I]{rpppl}. We also compared the AMPLE model with the 3GPP model, the ABG model, the CI model, and the models with simulation calibrations. The results showed that the AMPLE model outperforms these models by considering environments. 

Future works of the AMPLE model are mainly based on the discussions in \cite{ample00,ample01}, and the prediction ability of the AMPLE model under simple environments need further enhancement, such as adaptive parameter reducing, LOS simplification, and so on.

\appendix
We consider maximum likelihood with gradient descent to extract the parameters of the AMPLE model, the CI model, and the ABG model. The detailed extraction process for these three models are provided in this appendix. 
\subsection{The AMPLE Model}
\label{APP_AMPLE}
As shown in \eqref{WtdPL_LkTrm}, the AMPLE model can be computed as
\begin{equation}
	\text{PL}_z\,[\text{dB}] = A + \sum_{m=1}^{M} D_mn_m+p_zX+10 \gamma \log_{10}(f_z)+\Psi_{\sigma}.
	\tag{\ref{WtdPL_LkTrm}} 
\end{equation} 
With the random variable $\Psi_{\sigma}$, the path loss follows a normal distribution with $N[\mu(A,n_{m},X,\gamma),{\sigma}^{2}]$, where \cite{gus15}
\begin{equation}
	\begin{aligned}
	\label{MeanofWPL}
	\mu(A,n_{m},X,\gamma) = A + \sum_{m=1}^{M} &D_mn_m+p_zX\\ &+ 10 \gamma \log_{10}(f_z).
\end{aligned}
\end{equation} 
Following that, we first compute the PDF of \eqref{WtdPL_LkTrm}, which can be expressed as 
\begin{equation}
	\begin{aligned}
		\label{PDFPL}
		P(l_z;\mu(A,&n_{m}, X,\gamma),\sigma) =
		\\ &\frac{1}{\sqrt{2\pi}\sigma}\mathrm{exp}\bigg(-\frac{(l_z-\mu(A,n_{m},X,\gamma))^{2}}{2\sigma^{2}}\bigg),
	\end{aligned}
\end{equation}
where $l_z$ is the $z$th path loss value. Then, the joint PDF (i.e., likelihood function) can be expressed as  
\begin{equation}
	\begin{aligned}
		\label{Lfun}
		F(\mu(A,&n_{m},X,\gamma),\sigma) =
		\\ 
		& \prod_{z=1}^{Z}\frac{1}{\sqrt{2\pi}\sigma}\mathrm{exp}\bigg(-\frac{(l_z-\mu(A,n_{m},X,\gamma) )^{2}}{2\sigma^{2}}\bigg).
	\end{aligned}	
\end{equation}
By converting the likelihood function $F$ into its  negative natural logarithm form,  the problem of finding the optimal parameters can be formulated as minimizing this negative log-likelihood function, which can be expressed as
\begin{equation}
\label{minAMPLE}
\begin{aligned}
	\arg\min_{\substack{A,n_{m}, \\X, \gamma,\sigma}} (-\mathrm{ln} F(\mu(A,n_{m},X,\gamma),\sigma)).
\end{aligned}
\end{equation}
Since we use gradient descent to regress the parameters, the partial derivatives of $-\mathrm{ln} F(\mu(A,n_{m},X,\gamma),\sigma)$ with respect to all parameters are required, including PLEs $n_m$, intercept $A$, penetration loss $X$, frequency coefficient $\gamma$, and the standard deviation $\sigma$, which are
\begin{equation}
\begin{aligned}
	-\frac{\partial  \mathrm{ln}F(\mu(A,n_{m},X,\gamma),\sigma)}{\partial A} 
	&=  \\
	&\hspace{-0.2cm} \sum_{z=1}^{Z}  \bigg(\frac{\mu(A,n_{m},X,\gamma)-l_z}{\sigma^{2}} \bigg),  
\end{aligned} 
\label{PD_A}
\end{equation}
\begin{equation}
\begin{aligned}		
	-\frac{\partial  \mathrm{ln}F(\mu(A,n_{m},X,\gamma),\sigma)}{\partial n_m} 
	&=\\
	&\hspace{-0.85cm}\sum_{z=1}^{Z}  \bigg(\frac{(\mu(A,n_{m},X,\gamma)-l_z)D_m}{\sigma^{2}} \bigg),
\end{aligned}  
\label{PD_nm}	
\end{equation}
\begin{equation}	
\begin{aligned}		
-\frac{\partial  \mathrm{ln}F(\mu(A,n_{m},X,\gamma),\sigma)}{\partial X} 
	&= 	\\
	&\hspace{-0.5cm}\sum_{z=1}^{Z}  \bigg( \frac{(\mu(A,n_{m},X,\gamma)-l_z)p_z}{\sigma^{2}} \bigg),
\end{aligned} 
\label{PD_X}	
\end{equation}
\begin{equation}
\begin{aligned}
	-\frac{\partial  \mathrm{ln}F(\mu(A,n_{m},X,\gamma),\sigma)}{\partial \gamma} 
	&= \\
	&\hspace{-2cm}\sum_{z=1}^{Z}  \bigg( \frac{(\mu(A,n_{m},X,\gamma)-l_z)10\log_{10}(f_z)}{\sigma^{2}} \bigg),  
\end{aligned} 
\label{PD_f}	
\end{equation}
\begin{equation}
\begin{aligned}
	-\frac{\partial  \mathrm{ln}F(\mu(A,n_{m},X,\gamma),\sigma)}{\partial \sigma} 
	&= \\
	&\hspace{-1cm}\sum_{z=1}^{Z}  \bigg( \frac{1}{\sigma} - \frac{(l_z-\mu(A,n_{m},X,\gamma))^2}{\sigma^{3}} \bigg).   
\end{aligned} 
\label{PD_sigma} 
\end{equation}
Note that the initial values of gradient descent are chosen based on our previous works in \cite{ample00,ample01}, and the step size is set as $2\times10^{-6}$.

\subsection{The CI Model}
\label{APP_CI}

As shown in \eqref{PLCI1m}, the CI path loss model can be expressed as \cite{5gcmsig, rpppl}
\begin{equation}
	\begin{aligned}
		\text{PL}_z^{\text{CI}}(f_z,d_z)\,[\text{dB}] &= 20\log_{10}\left(\frac{4\pi f_z \times 10^9}{c}\right)
		\\
		& + 10n\log_{10}\left(\frac{d_z}{1\,\text{m}}\right) + \Psi_{\sigma}^{\text{CI}}.
	\end{aligned}	
	\tag{\ref{PLCI1m}}
\end{equation}
With random variable $\Psi_{\sigma}^{\text{CI}}$, the mean of the CI model is 
\begin{equation}
		\label{MeanofCI}
		\mu(n) = 20\log_{10}\left(\frac{4\pi f_z \times 10^9}{c}\right) + 10n\log_{10}(d_z).
\end{equation}
Following that, the PDF of the CI model is
\begin{equation}
	\begin{aligned}
		\label{PDFCI}
		P(l_z;\mu(n),\sigma)=\frac{1}{\sqrt{2\pi}\sigma}\mathrm{exp}\bigg(-\frac{(l_z-\mu(n))^{2}}{2\sigma^{2}}\bigg).
	\end{aligned}
\end{equation}
By combining all data points, the likelihood function is written as
\begin{equation}
	\begin{aligned}
		\label{LfunCI}
		F(\mu(n),\sigma) = \prod_{z=1}^{Z}\frac{1}{\sqrt{2\pi}\sigma}\mathrm{exp}\bigg(-\frac{(l_z-\mu(n) )^{2}}{2\sigma^{2}}\bigg).
	\end{aligned}	
\end{equation}
Similarly, the parameters within the CI model can be treated as minimizing the negative natural logarithm of \eqref{LfunCI}, which is
\begin{equation}
	\label{minCI}
	\begin{aligned}
		\arg\min_{\substack{n,\sigma}} (-\mathrm{ln} F(\mu(n),\sigma)).
	\end{aligned}
\end{equation}
Therefore, the partial derivatives of $-\mathrm{ln} F(\mu(n),\sigma)$ including CI PLE $n$ and standard deviation $\sigma$ can be computed as
\begin{equation}
	\begin{aligned}		
		-\frac{\partial  \mathrm{ln}F(\mu(n),\sigma)}{\partial n} =\sum_{z=1}^{Z}  \bigg(\frac{(\mu(n)-l_z)10\log_{10}(d_z)}{\sigma^{2}} \bigg),
	\end{aligned}  
	\label{PDCI_n}	
\end{equation}
\begin{equation}
	\begin{aligned}
		-\frac{\partial  \mathrm{ln}F(\mu(n),\sigma)}{\partial \sigma} =\sum_{z=1}^{Z}  \bigg( \frac{1}{\sigma} - \frac{(l_z-\mu(n))^2}{\sigma^{3}} \bigg).   
	\end{aligned} 
	\label{PDCI_sigma} 
\end{equation}
We choose the initial values of both $n$ and $\sigma$ based on \cite[Table 6]{5gcmsig}, and the step size of gradient descent is set as $2\times10^{-6}$.

\subsection{The ABG Model}
\label{APP_ABG}

As shown in \eqref{PLABG1m}, the ABG path loss model can be computed as \cite{5gcmsig, rpppl}
\begin{equation}
	\begin{aligned}
		\text{PL}_z^{\text{ABG}}(f_z,d_z)\,[\text{dB}] &= 10\alpha\log_{10}\left(\frac{d_z}{1\,\text{m}}\right) \\ 
		&+\beta + 10\gamma^{\text{ABG}}\log_{10}\left(\frac{f_z}{1\,\text{GHz}}\right)+ \Psi_{\sigma}^{\text{ABG}}.
	\end{aligned}
		\tag{\ref{PLABG1m}}	
\end{equation}
With random variable $\Psi_{\sigma}^{\text{ABG}}$, the mean of the ABG model is
\begin{equation}
	\begin{aligned}
	\label{MeanofABG}
	\mu(\alpha,\beta,\gamma^{\text{ABG}}) = 10\alpha\log_{10}&\left(\frac{d_z}{1\,\text{m}}\right) 
	+\beta  \\& +10\gamma^{\text{ABG}}\log_{10}\left(\frac{f_z}{1\,\text{GHz}}\right).
	\end{aligned}
\end{equation}
Following that, the PDF of the ABG model is
\begin{equation}
	\begin{aligned}
		\label{PDFABG}
		P(l_z;\mu(\alpha,\beta,\gamma^{\text{ABG}}),\sigma) &= \\&\hspace{-.5cm}\frac{1}{\sqrt{2\pi}\sigma}\mathrm{exp}\bigg(-\frac{(l_z-\mu(\alpha,\beta,\gamma^{\text{ABG}}))^{2}}{2\sigma^{2}}\bigg),
	\end{aligned}
\end{equation}
Along with all data points, the likelihood function can be expressed as
\begin{equation}
	\begin{aligned}
		\label{LfunABG}
		F(\mu(\alpha,\beta,\gamma^{\text{ABG}}),&\sigma) = \\&\prod_{z=1}^{Z}\frac{1}{\sqrt{2\pi}\sigma}\mathrm{exp}\bigg(-\frac{(l_z-\mu(\alpha,\beta,\gamma^{\text{ABG}}) )^{2}}{2\sigma^{2}}\bigg).
	\end{aligned}	
\end{equation}
Similarly, the parameters within the ABG model can be formulated as minimizing the negative natural logarithm of \eqref{LfunABG}, and can be expressed as
\begin{equation}
	\label{minABG}
	\begin{aligned}
		\arg\min_{\substack{\alpha,\beta,\gamma^{\text{ABG}},\sigma}} (-\mathrm{ln} F(\mu(\alpha,\beta,\gamma^{\text{ABG}}),\sigma)).
	\end{aligned}
\end{equation}
In the meantime, the partial derivatives of $-\mathrm{ln} F(\mu(\alpha,\beta,\gamma^{\text{ABG}}),\sigma)$ including PLE $\alpha$, intercept $\beta$, frequency coefficient $\gamma^{\text{ABG}}$, and the standard deviation $\sigma$ can be computed as
\begin{equation}
	\begin{aligned}		
		-\frac{\partial  \mathrm{ln}F(\mu(\alpha,\beta,\gamma^{\text{ABG}}),\sigma)}{\partial \alpha}& = \\ &\hspace{-1.5cm}\sum_{z=1}^{Z}  \bigg(\frac{(\mu(\alpha,\beta,\gamma^{\text{ABG}})-l_z)10\log_{10}(d_z)}{\sigma^{2}} \bigg),
	\end{aligned}  
	\label{PDABG_A}	
\end{equation}
\begin{equation}
	\begin{aligned}		
		-\frac{\partial  \mathrm{ln}F(\mu(\alpha,\beta,\gamma^{\text{ABG}}),\sigma)}{\partial \beta} =\sum_{z=1}^{Z}  \bigg(\frac{(\mu(\alpha,\beta,\gamma^{\text{ABG}})-l_z)}{\sigma^{2}} \bigg),
	\end{aligned}  
	\label{PDABG_B}	
\end{equation}
\begin{equation}
	\begin{aligned}		
		-\frac{\partial  \mathrm{ln}F(\mu(\alpha,\beta,\gamma^{\text{ABG}}),\sigma)}{\partial \gamma^{\text{ABG}}} &= \\ & \hspace{-1.5cm}\sum_{z=1}^{Z}  \bigg(\frac{(\mu(\alpha,\beta,\gamma^{\text{ABG}})-l_z)10\log_{10}(f_z)}{\sigma^{2}} \bigg),
	\end{aligned}  
	\label{PDABG_G}	
\end{equation}
\begin{equation}
	\begin{aligned}
		-\frac{\partial  \mathrm{ln}F(\mu(\alpha,\beta,\gamma^{\text{ABG}}),\sigma)}{\partial \sigma} &= \\ & \hspace{-.3cm}\sum_{z=1}^{Z}  \bigg( \frac{1}{\sigma} - \frac{(l_z-\mu(\alpha,\beta,\gamma^{\text{ABG}}))^2}{\sigma^{3}} \bigg).   
	\end{aligned} 
	\label{PDABG_sigma} 
\end{equation}
For LOS and NLOS environments, the initial values are based on \cite[Table III]{rpppl} and \cite[Table 6]{5gcmsig}, respectively. The step size of gradient descent is set as $2\times10^{-6}$.

\vfill

\end{document}